\documentclass[namedreferences]{solarphysics}

\usepackage[figuresright]{rotating}
\usepackage[hyperref,optionalrh,solaromanenum]{spr-sola-addons}
\usepackage{graphicx}
\usepackage{txfonts}
\usepackage{natbib}
\usepackage{amssymb}
\usepackage{color}
\usepackage{booktabs}

\usepackage{caption}
\DeclareCaptionLabelSeparator{myspace}{\rule{4pt}{0pt}}
\hypersetup{
    final=true,
    pageanchor=true,
    colorlinks=true,
    breaklinks=true,
    linkcolor=blue,
    citecolor=blue,
    urlcolor=blue,
    pdfpagemode=UseNone,
    pdftitle={Chromospheric Synoptic Maps of Polar Crown Filaments},
    pdfauthor={Diercke and Denker},
    pdfsubject={solar physics},
    pdfkeywords={Chromosphere, Quiet; 
        Prominences, Quiescent; 
        Prominences, Magnetic Field;
        Solar Cycle, Observations;
        Instrumentation and Data Management}}

\newcommand\phn{\phantom{0}}
\newcommand\degr{\ensuremath{^\circ}}

\begin{document}

\begin{article}

%
%

\begin{opening}

%
%

\title{Chromospheric Synoptic Maps of Polar Crown Filaments}

\author[addressref={aff1,aff2},corref,email={adiercke@aip.de}]
    {\inits{A.\ }\fnm{A.\ }\lnm{Diercke}\orcid{0000-0002-9858-0490}}
\author[addressref=aff1, corref]
    {\inits{C.\ }\fnm{C.\ }\lnm{Denker}\orcid{0000-0003-1054-766X}} 

\address[id={aff1}]{Leibniz-Institut f\"ur Astrophysik Potsdam (AIP),
    An der Sternwarte 16, 14482 Potsdam, Germany}
\address[id={aff2}]{Universit\"at Potsdam, Institut f\"ur Physik und 
    Astronomie, Karl-Liebknecht-Stra\ss{}e 24/25, 14476 Potsdam, Germany}

\runningauthor{A.\ Diercke and C.\ Denker}
\runningtitle{Chromospheric Synoptic Maps of Polar Crown Filaments}

%
%

\begin{abstract}
Polar crown filaments form above the polarity inversion line between the old magnetic flux of the previous cycle and the new magnetic flux of the current cycle. Studying their appearance and their properties can lead to a better understanding of the solar cycle. We use full-disk data of the \textit{Chromospheric Telescope} (ChroTel) at the \textit{Observatorio del Teide}, Tenerife, Spain, which were taken in three different chromospheric absorption lines (H$\alpha$ $\lambda$6563\,\AA, \mbox{Ca\,\textsc{ii}\,K} $\lambda$3933\,\AA, and \mbox{He\,\textsc{i}}~$\lambda$10830\,\AA), and we create synoptic maps. In addition, the spectroscopic \mbox{He\,\textsc{i}} data allow us to compute Doppler velocities and to create synoptic Doppler maps. ChroTel data cover the rising and decaying phase of Solar Cycle 24 on about 1000 days between 2012 and 2018. Based on these data, we automatically extract polar crown filaments with image-processing tools and study their properties. We compare contrast maps of polar crown filaments with those of quiet-Sun filaments. Furthermore, we present a super-synoptic map summarizing the entire ChroTel database. In summary, we provide statistical properties, \textit{i.e.} number and location of filaments, area, and tilt angle for both the maximum and declining phase of Solar Cycle~24. This demonstrates that ChroTel provides a promising data set to study the solar cycle. 
\end{abstract}

\end{opening}

%
%

\section{Introduction}\label{s:intro} 

Polar crown filaments (PCFs) form at the intersection of old flux from the previous cycle in the polar region and new flux of the present cycle, which is transported out of the activity belts toward the Poles. The flux of opposite polarity builds a neutral line above which a filament channel can form \citep{Martin1998a, Mackay2010}. The PCFs appear inside the filament channel. They are usually long-lived structures and can last over several days up to weeks. The PCFs appear shortly after the magnetic-flux reversal around the maximum of the solar cycle at mid-latitudes. During the course of the solar cycle, PCFs form closer to the Pole as the neutral line shifts closer to the Pole. Around the time of magnetic polarity reversal, PCFs disappear from the solar surface \citep{Leroy1983, Leroy1984}. This cyclic behavior is known since the beginning of the $20$th century \citep{Cliver2014}. However, studies covering the current cycle are still rare \citep{Xu2018}. This propagation toward the Pole throughout the cycle is known as the ``dash-to-the-Pole'' \citep{Evershed1917, Cliver2014}. 

The PCFs belong to the category of quiet-Sun filaments, which appear outside of the activity belts. Another class of filaments are active-region filaments, which reside in the vicinity of active regions. Filaments that cannot be assigned to either active-region or quiet-Sun filaments are called intermediate filaments \citep{Martin1998a}. 

To study the cyclic behavior of PCFs, a more global view of the Sun is needed for which regular long-term observations of the solar chromosphere are necessary. From these observations, synoptic maps can be produced, which allow us to study large-scale patterns over a longer periods of time. \citet{Carrington1858} first created synoptic maps of sunspot observations from which a cyclic behavior could be determined, \textit{i.e.} most prominently Sp\"orer's Law \citep{Cameron2017}. The variety of synoptic maps nowadays increased in the last 150~years, and many more physical parameters are available to be studied. One well-known example of synoptic maps is the hand-drawn McIntosh maps \citep{McIntosh2005, Gibson2017}, which allow us to determine a relation between open magnetic structures (coronal holes) and closed magnetic structures (filaments or active regions). Another example are long-term studies of filaments with full-disk images from the \textit{Kodaikanal Observatory} (KO) of the Indian Institute of Astrophysics (IAA) from 1914\,--\,2007 \citep{Chatterjee2017}. The work of \citet{Hao2015} shows a detailed study of filaments and their statistical properties for the years 1988\,--\,2013 for H$\alpha$ observations of the \textit{Big Bear Solar Observatory} (BBSO) in California. Finally, \citet{Xu2018} extracted the positions of PCFs for four cycles (1973 -- 2018) from H$\alpha$ observations at BBSO and at the \textit{Kanzelh\"ohe Solar Observatory} (KSO) in Austria, which are directly relevant to the present work.

Morphological image processing is a powerful tool for automatically analyzing  images and is widely used in solar physics. \citet{Robbrecht2004} carried out a study on the automatic recognition of coronal mass ejections (CME) by using the Hough transform \citep{Gonzalez2002}, and in addition, they used morphological closing to separate different CMEs. Another application is soft morphological filters to reduce noise in images as presented by \citet{Marshall2006} for the reduction of cosmic-ray induced noise in images of the \textit{Large Angle Spectroscopic Coronagraph} \citep[LASCO:][]{Brueckner1995}. Morphological image processing using the blob analysis \citep{Fanning2011}, to select connected regions in an image, enabled for example a statistical analysis of pores \citep{Verma2014} and of historical sunspot drawings of the years 1861--1894 from Sp\"orer \citep{Diercke2015}. \citet{Qu2005} used morphological image processing for automatic threshold extraction to select dark objects in H$\alpha$ line-core images based on a Support Vector Machine \citep[SVM:][]{Cortes1995} to identify the spines and footpoints of filaments and to recognize the disappearance of filaments.

In Section~\ref{s:obs}, we give an overview of the database of the \textit{Chromospheric Telescope} \citep[ChroTel: ][]{Kentischer2008, Bethge2011}, and we present synoptic maps of three different chromospheric spectral lines, which will be used to study the poleward migration of high-latitude filaments during Solar Cycle 24, as well as the \mbox{He\,\textsc{i}} $\lambda$10830\,\AA\ Doppler velocity. We introduce the data handling and a reduction method for the ChroTel data, as well as a method to extract filaments from the synoptic maps based on morphological image processing (Section~\ref{s:method}). Basic physical parameters, \textit{i.e.} area, tilt angle, and location on the disk, are extracted automatically with morphological image processing tools (Section~\ref{s:results}). We separated the high-latitude filaments into different categories, \textit{i.e.} quiet-Sun filaments and PCFs, where we study their different appearances and  compare their intensity variations in H$\alpha$ $\lambda$6563\,\AA\ and \mbox{He\,\textsc{i}}~$\lambda$10830\,\AA. The categorization of different objects on the Sun will be discussed in the context of future labeling approaches with neural networks \citep{Goodfellow2016}.

%
\section{Observations}\label{s:obs} 
%

\begin{figure} 
\centering
\includegraphics[width=1.0\textwidth]{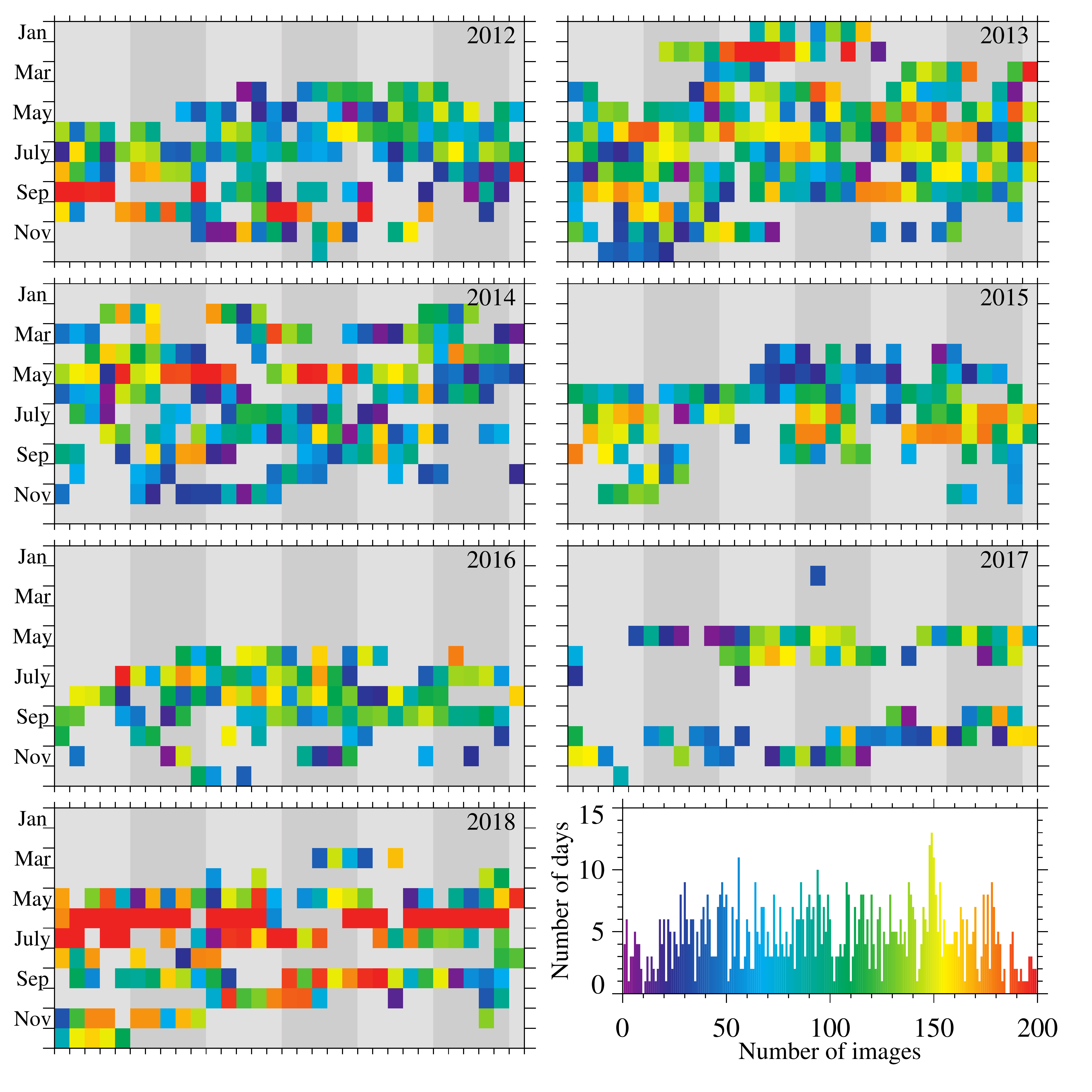}
\caption{Overview of all data between 2012 and 2018. The number of images per day is color-coded. The \textit{gray-shaded areas} indicate a time interval of five days during a given month. The \textit{histogram in the lower right corner} indicates the number of images for all days between 1\,--\,199 images per day. More than 200 images per day were recorded on 46 days, which are indicated in \textit{red}.} 
\label{fig:overview}
\end{figure}

ChroTel is mounted two levels above ground on a projecting roof of the \textit{Vacuum Tower Telescope} \citep[VTT: ][]{vonderLuehe1998} at the \textit{Observatorio del Teide} on Tenerife, Spain. It is a ten-cm aperture full-disk telescope operating automatically since 2011. It records chromospheric images with $2048\,\times\,2048$~pixels using three Lyot-type birefringent narrow-band filters in H$\alpha$ $\lambda$6562.8\,\AA\ with a FWHM of 0.5\,\AA, \mbox{Ca\,\textsc{ii}\,K} at $\lambda$3933.7\,\AA\ with a FWHM of 0.3\,\AA, and \mbox{He\,\textsc{i}} $\lambda$10830\,\AA\ with a FWHM of 0.14\,\AA. In the \mbox{He\,\textsc{i}} line, the instrument scans seven non-equidistant filter positions within $\pm$3\,\AA\ around the line core of the red component, whereby liquid-crystal variable retarders allow a rapid adjustment of the seven wavelength points \citep{Bethge2011}. The quantum efficiency is highest for H$\alpha$ at 63\,\% and reduces for \mbox{Ca\,\textsc{ii}\,K} to 39\,\% and for \mbox{He\,\textsc{i}} to 2\,\% \citep{Bethge2011}. The exposure time in H$\alpha$ is 100\,ms. Because of the low quantum efficiency of the other two channels, the exposure time was selected to be higher with 300\,ms for each filter position around the  \mbox{He\,\textsc{i}} line and 1000\,ms for \mbox{Ca\,\textsc{ii}\,K}.

ChroTel contributes to a variety of scientific topics \citep{Kentischer2008, Bethge2011}, \textit{e.g.} statistical studies of filament properties, supersonic downflows, and chromospheric sources of the fast solar wind, among others. For the last two topics, the calculation of Doppler velocities from the spectroscopic \mbox{He\,\textsc{i}} data is required. In addition, ChroTel full-disk images monitor large-scale structures and give an overview of chromospheric structures in the vicinity of solar eruptive events. The cadence between images of the same filter is three minutes in the standard observing mode. Higher cadences of up to ten seconds for H$\alpha$ and \mbox{Ca\,\textsc{ii}\,K} and 30 seconds for \mbox{He\,\textsc{i}} are possible with single-channel observations \citep{Bethge2011}.

Between 2012 and 2018, ChroTel took H$\alpha$ data of the Sun on 974 days. For each day, we downloaded all recorded images but kept only the best image of the day. The images were selected by calculating the \textsf{Median Filter-Gradient Similarity} \citep[MFGS:][]{Deng2015, Denker2018}, which computes the similarity of the gradients of the original images and those of median-filtered images. Along with the H$\alpha$ images, we downloaded the images closest in time for \mbox{Ca\,\textsc{ii}\,}K and \mbox{He\,\textsc{i}}. There are some days when ChroTel observations covered only H$\alpha$ but with a higher cadence of one minute.

The total amount of data in the ChroTel archive is visualized in Figure~\ref{fig:overview}. We display in rainbow colors the number of images in H$\alpha$ for each day in the year. Usually, observations commence in April and last until November. In some years, the observations started already in January and continued until December. In 2017, less data were taken because of technical problems. The regular automatic observations started again in mid-2018. In that year, some days contained observations with more than 200 images, which implies time-series of more then ten hours. The ChroTel archive is hosted by the Leibniz Institute for Solar Physics (KIS) in Freiburg, Germany (\href{ftp://archive.leibniz-kis.de/archive/chrotel}{ftp://archive.leibniz-kis.de/archive/chrotel}). Table~\ref{tbl:numimg} gives an overview of the total number of images in the ChroTel archive for each spectral line.

%
\begin{table}
\caption{Summary of ChroTel observations in different spectral lines 
    between 2012 and 2018}\label{tbl:numimg}
\begin{tabular}{lcc}     
\hline
Spectral line & Days of observations & Number of images\rule[-2pt]{0pt}{10pt} \\
\hline
H$\alpha$                 & 974 &     111,633 \\
\mbox{He\,\textsc{i}}     & 916 & \phn 94,670 \\
\mbox{Ca\,\textsc{ii}\,K} & 899 & \phn 90,353 \\
\hline
\end{tabular}
\end{table}

%
\section{Methods}\label{s:method} 
%

Level 1.0 data, which are available in the ChroTel archive, went through the standard dark- and flat-field correction and were converted to the format of the Flexible Image Transport System \citep[FITS:][]{Wells1981, Hanisch2001}. In preprocessing, we rotated and rescaled the images, so that the radius corresponds to $r=1000$~pixels, which yields an image scale of about 0.96$^{\prime\prime}$ pixel$^{-1}$. A constant radius rather than a constant image scale makes it easier to construct synoptic maps in an automated way. The images were corrected for hot pixels by removing these strong intensity spikes from the image by replacing them with the local median value using routines from the \textsf{sTools} library \citep{Kuckein2017IAU}. We noticed that in images that were recorded early in the morning and late in the afternoon, the solar disk appeared deformed and had the form of an ellipse. This is caused by differential refraction because of significantly varying air mass across the solar disk. Thus, we fitted an ellipse to the solar disk and stretched the solar disk to the theoretical value of the solar radius recorded in the FITS header. For the \mbox{Ca\,\textsc{ii}\,K} images, we observed that the solar disk was not centered in the images such that in these images a geometrical distortion is visible at the limb. Furthermore, all images were corrected for limb-darkening by fitting a fourth-order polynomial to the intensity profile. In a second step, bright areas and dark sunspots were excluded from the intensity profile and the polynomial fit was repeated \citep[\textit{e.g.} ][]{Diercke2018, Denker1999}. The images were corrected by dividing the images with the resulting  center-to-limb-variation profile. Because of intensity variations introduced by non-uniform filter transmission, we corrected all images with a recently developed method by \citet{Shen2018} using Zernike polynomials to yield an even background by dividing the images by their intensity-variation profile. First, we created a mask, where we excluded bright and dark areas. The resulting intensity profile was fitted with a linear combination of 36 Zernike polynomials. The resulting linear least-square problem was solved with singular value decomposition. From the back-substitution, we obtained 36 coefficients for Zernike polynomials.

\begin{figure}[t] 
\centering
\includegraphics[width=1.0\textwidth]{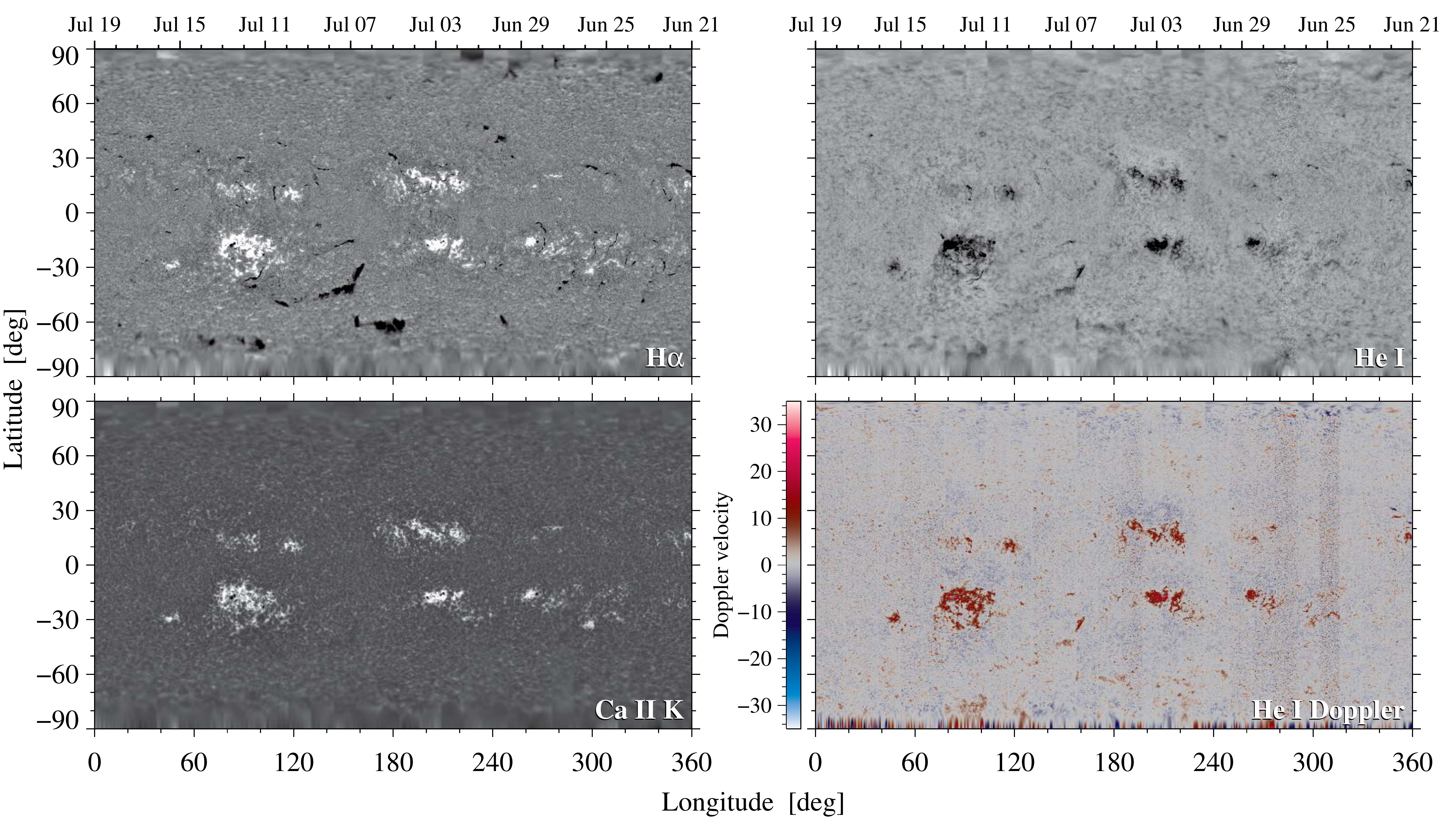}
\caption{Synoptic line-core intensity maps of the strong chromospheric absorption lines H$\alpha$ $\lambda$6563\,\AA, \mbox{Ca\,\textsc{ii}\,K} $\lambda$3933\,\AA, and \mbox{He\,\textsc{i}}~$\lambda$10830\,\AA\ including a \mbox{He\,\textsc{i}} Doppler map (\textit{top-left to bottom-right}), which cover Carrington rotation No.~2125 from 21~June to 19~July~2012. The Doppler velocities were derived with the method explained by \citet{Shen2018} and were scaled between $\pm35$\,km\,s$^{-1}$, where red and blue indicate down- and upflows, respectively.}
\label{fig:synoptic}
\end{figure}

To create the Carrington maps, one image per day was rotated to the corresponding longitudinal positions on a Carrington grid using the mapping routines of \textsf{SolarSoftWare} \citep[SSW:][]{Freeland1998, Bentely1998} for differential rotation correction. These image slices were merged creating a Carrington map for each solar rotation with a longitudinal sampling of 0.1$^\circ$. The image slices of one day were overlapping with the previous and following day by two hours. If an image of the previous or following day were missing, the input image was rotated that it covered in total 48 hours of solar rotation. Longer gaps were not covered with data and were left blank. Finally, the grid was transferred to heliographic coordinates and converted to an equidistant grid, which caused noticeable stretching at the polar regions. The process was repeated for all three chromospheric lines (Figure~\ref{fig:synoptic}). In the case of the \mbox{He\,\textsc{i}} filtergrams, we created a synoptic map individually for each filtergram. As a result, we can produce synoptic Doppler maps (Figure~\ref{fig:synoptic}d) by subtracting the two line-wing filtergrams as described by \citet{Shen2018}. In the H$\alpha$ data (Figure~\ref{fig:synoptic}a), we clearly recognize the filaments as elongated black structures, whereby the active-region plage areas appear bright. The sample map of Carrington rotation 2125 also contains PCFs (Figure~\ref{fig:synoptic}). In the \mbox{He\,\textsc{i}} data, active regions and filaments appear dark. The chromospheric emission of the \mbox{Ca\,\textsc{ii}\,K} line is greatly enhanced in locations, where the magnetic field is strong, \textit{i.e}, in plage regions and the chromospheric network. The filaments are, however, not recognizable in the line core of the \mbox{Ca\,\textsc{ii}\,K} line. In the Doppler maps, plage regions exhibit downflows, whereas regions with stronger upflows are located in the surroundings.

\begin{figure}[t] 
\centering
\includegraphics[width=1.0\textwidth]{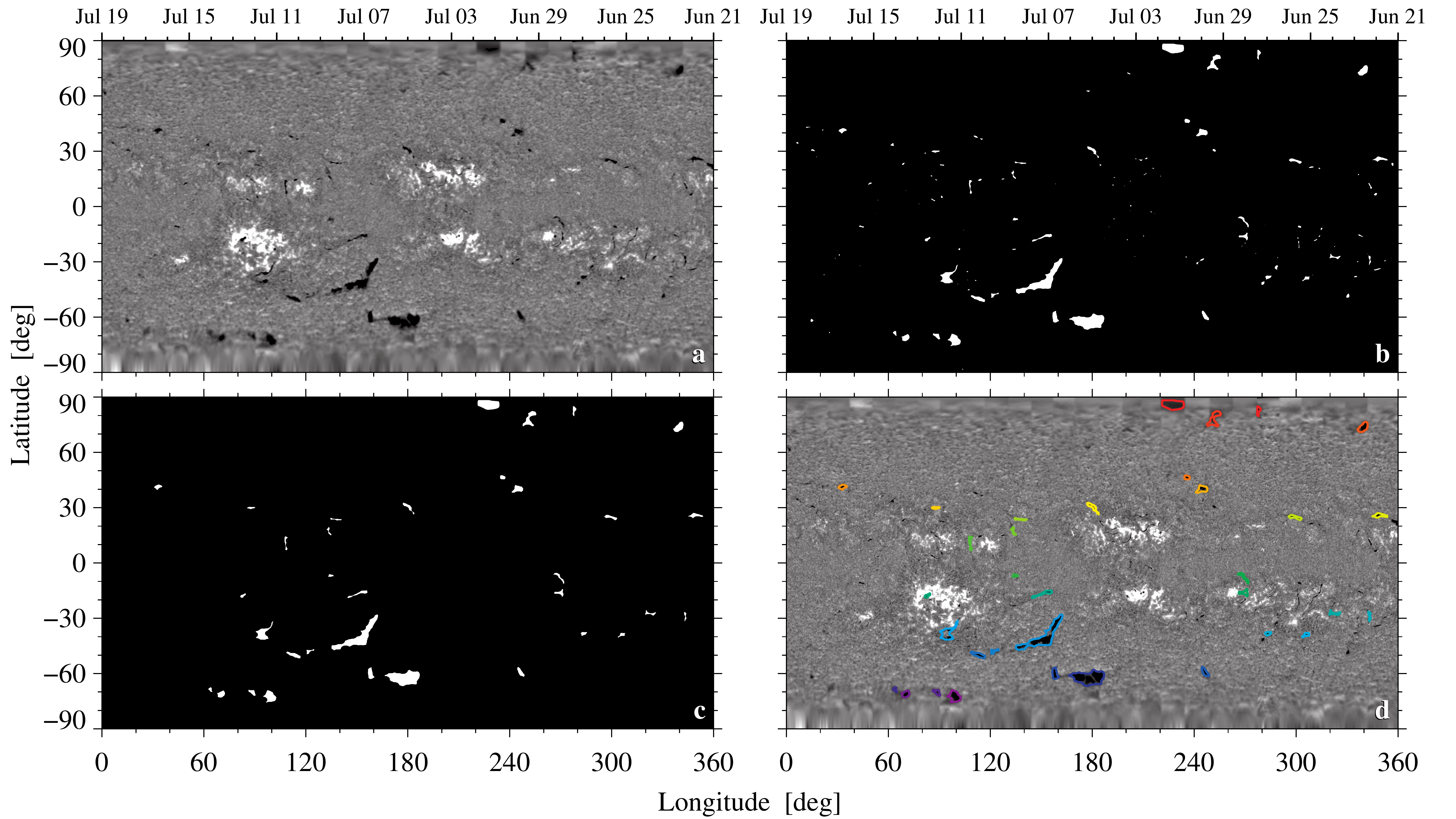}
\caption{Morphological filament extraction from an H$\alpha$ synoptic map. \textbf{(a)} H$\alpha$ synoptic map after application of the Perona--Malik-filter. \textbf{(b)} Mask of dark structures after morphological closing. \textbf{(c)} Mask after removal of small dark regions. \textbf{(d)} Resulting map with color-coded contours of extracted filaments, indicating the different detected objects.}
\label{fig:morph}
\end{figure}

The synoptic ChroTel maps contain, among others, information about the number, area, location, and orientation of filaments. We apply basic morphological image processing tools to H$\alpha$ maps, which will be outlined in the following, to extract this information for further analysis. Figure~\ref{fig:morph} shows as an example the procedure for Carrington rotation 2125, which was used to extract the filaments. The background of the images was perfectly flat because the non-uniform background was removed by dividing individual full-disk images with the intensity variation expressed by Zernike polynomials. The images were normalized with respect to the quiet-Sun intensity. We used the Perona--Malik filter \citep{Perona1990}, an anisotropic diffusion algorithm, to smooth structures while preserving the borders of elongated dark structures such as filaments (Figure~\ref{fig:morph}a). The filter was used with the gradient-modulus threshold controlling the diffusion $\kappa$ = 30 and with 15 iterations in the exponential implementation, which favors high-contrast over low-contrast structures. Thus, we used a global threshold to extract dark structures, which was selected by visual inspection. The thresholds were 0.3 and 0.9 times the average quiet-Sun intensity. In addition, we applied morphological closing with a circular kernel with a diameter of 56\,pixels to close small gaps between connected structures (Figure~\ref{fig:morph}b). We used a blob analyzing algorithm \citep{Fanning2011}, which selects connected regions called ``blobs'' in the image based on the four-adjacency criterion, and applied it to a binary mask containing pixels belonging to dark structures. Because we only want to analyze large-scale PCFs and quiet-Sun filaments, dark structures with a size of 500~pixels or smaller, which include sunspots and small-scale active region filaments, were excluded from further analysis (Figure~\ref{fig:morph}c). The remaining structures are shown in Figure~\ref{fig:morph}d.

Several properties follow from the blob analysis such as the location of the detected filaments and their area in pixels. Other properties of filaments can be determined from fitting an ellipse to the structure, such as the tilt-angle, as well as semi-major and semi-minor axis, which gives a first approximation of the length and width of the filament. The location of filaments enables us to study the cyclic behavior of PCFs.

We focus in the following analysis on PCFs and other high-latitude filaments. We categorize PCFs as high-latitude filaments if their center-of-gravity is located at latitudes above/below $+50\degr$/$-50\degr$, as in the work of \citet{Hao2015}. Because of strong geometric distortions in synoptic maps near the Poles, structures above/below $+85\degr$/$-85\degr$ are excluded. Additional quiet-Sun filaments are selected if their center-of-gravity is located between $\pm30\degr$ and $\pm50\degr$.

\begin{figure} 
\centerline{
\includegraphics[width=1.\textwidth]{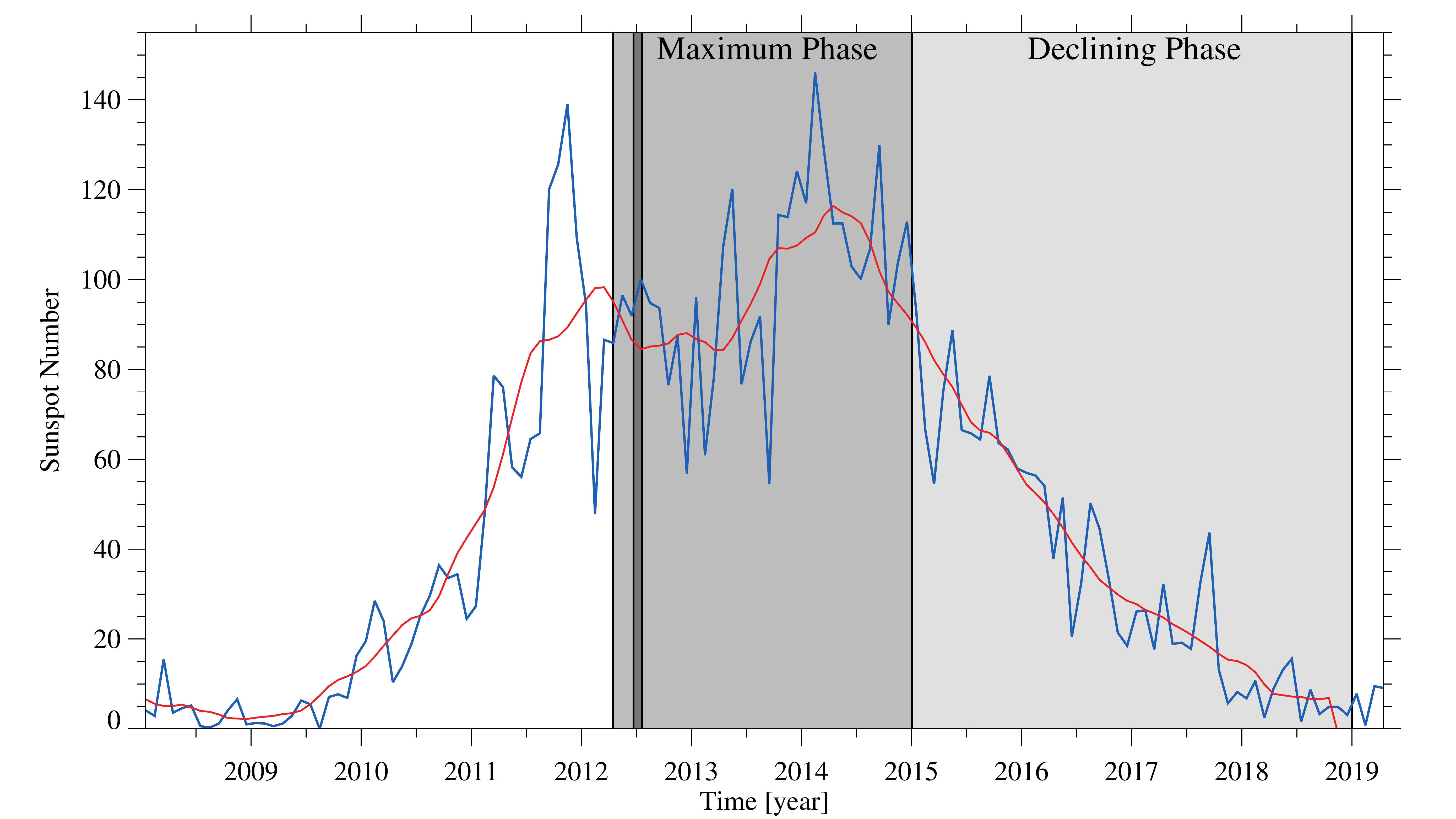}}
\caption{Monthly mean total sunspot number (\textit{blue}) and 13-month smoothed monthly total sunspot number (\textit{red}) for Solar Cycle 24 \citep{sidc}. The \textit{gray areas} indicate the two observing periods in which the ChroTel data were divided. The maximum phase starts on 13~April~2012 and lasts until the end of 2014. The declining phase covers the years 2015\,--\,2018. The \textit{small dark gray region} indicates Carrington rotation 2125, which is displayed in Figure~\ref{fig:synoptic}.}
\label{fig:tmsn}
\end{figure}

%
\section{Results}\label{s:results} 
%

The current Solar Cycle~24 (Figure~\ref{fig:tmsn}, the sunspot data for Solar Cycle~24 were provided by \citet{sidc}, \href{http://www.sidc.be/SILSO/}{www.sidc.be/SILSO/}) started in late 2008. In 2011 and 2014 two maxima were reached. \citet{Sun2015} date the polar magnetic-field reversal for the northern and southern hemisphere in November 2012 and March 2014, respectively. The minimum is expected by the end of 2019 or the beginning of 2020. The regular ChroTel observations started 2012, after the first maximum in 2011. They cover the second maximum in 2014 with a large number of observing days in that year, and monitor the decreasing phase of the cycle between 2015 and 2018. In the following analysis, we split the data set in two parts: the maximum phase contains the observations around the maximum between 2012\,--\,2014 (dark gray in Figure~\ref{fig:tmsn}, 528 observing days). The declining phase contains the observations between 2015\,--\,2018 after the magnetic polarity reversal (light gray in Figure~\ref{fig:tmsn}, 434 observing days).

\subsection{Global Distribution of Filaments in Latitude}

\begin{figure} 
\centerline{
\includegraphics[width=1.0\textwidth]{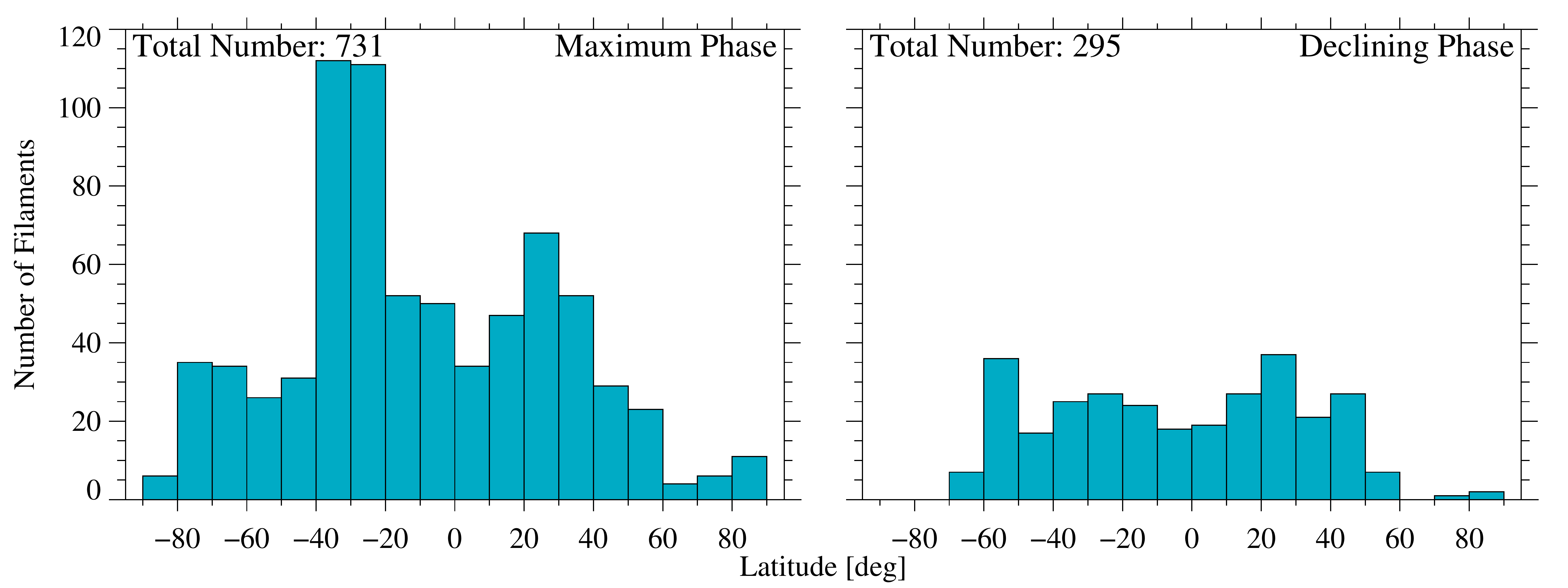}}
\caption{Distribution of detected filaments per latitude bin for the maximum (\textit{left}) and the declining (\textit{right}) phase. Each bin has a size of 10\degr.}
\label{fig:hnumber}
\end{figure}

The number of detected structures in each latitude bin is displayed in Figure~\ref{fig:hnumber} for both the maximum and declining phase. For the maximum phase, most detected filaments are located in the mid-latitudes between $\pm20\degr$ and $\pm50$\degr, where intermediate and quiet-Sun filaments are located. Fewer filaments are detected in the higher latitudes above/below $+60\degr$ and $-60\degr$, where more filaments appear in the southern hemisphere. This can be explained with the delayed polar magnetic field reversal in the southern hemisphere. In the declining phase, fewer filaments are detected because of lower solar activity but also because of the sparser coverage of ChroTel data. Nonetheless, we recognize fewer detected filaments at high latitudes, which is expected following the polar magnetic field reversal in both hemispheres.

\subsection{Latitude Dependence of Filaments across the Solar Cycle}

\begin{figure} 
\centering
\includegraphics[width=1.0\textwidth]{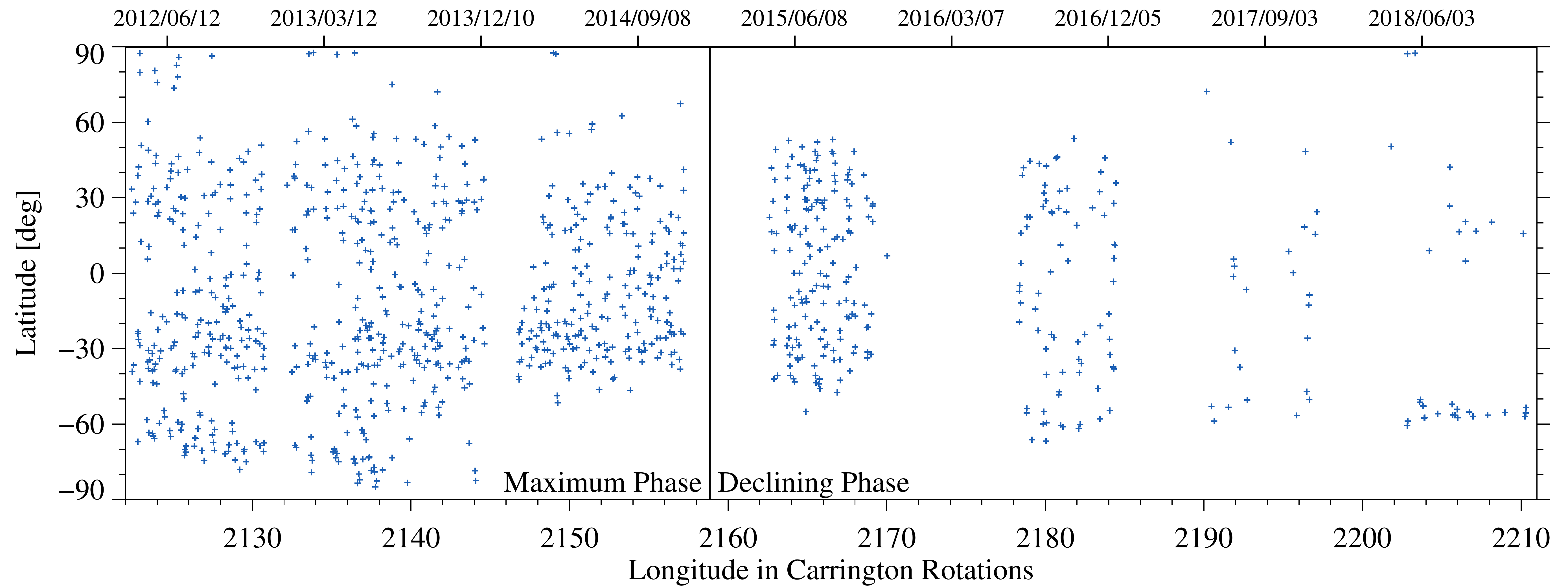}
\caption{Position of identified filaments for 90 Carrington rotation periods between 2012 and 2018. The line indicates the split of the data set in maximum and declining phase. The maximum phase includes the years 2012\,--\,2014 and the declining phase includes the years 2015\,--\,2018.} 
\label{fig:supersynopt}
\end{figure}

\begin{figure} 
\centerline{
\includegraphics[width=1.0\textwidth]{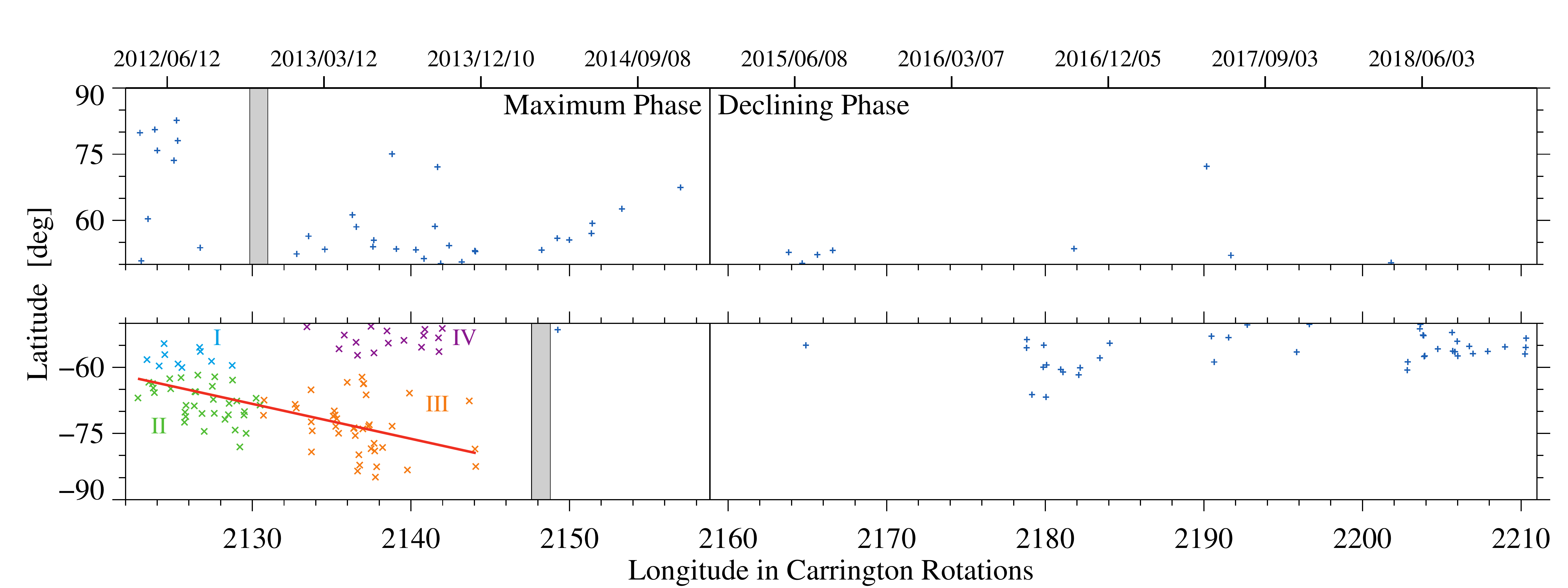}}
\caption{Butterfly diagram of high-latitude filaments above/below $+50\degr$/$-50\degr$. The \textit{gray areas} indicate the time of the polar magnetic field reversal in November 2012 for the northern hemisphere (\textit{top}) and March 2014 for the southern hemisphere (\textit{bottom}). The filaments of the southern hemisphere of 2012 and 2013 are divided in four clusters. The filaments belonging to Cluster~I (\textit{blue crosses}), Cluster~II (\textit{green crosses}), and Cluster~III (\textit{orange crosses}) were used to calculate the migration rate. The red line indicates the migration of polar crown filaments toward the Pole. Filaments belonging to Cluster~IV (\textit{violet crosses}) are quiet-Sun filaments and were not used in the computation of the migration rate. The \textit{vertical black} line indicates the the split of the data set into maximum and declining phase.}
\label{fig:butter30}
\end{figure}

We derived the location of the center-of-gravity for each detected object using blob analysis. In Figure~\ref{fig:supersynopt}, we display the location with respect to the corresponding Carrington rotation for the entire data set. The black line indicates the split into the maximum phase on the left and the declining phase on the right. In the maximum phase, we have more observations and therefore more detected filaments. Higher latitudes are more interesting for this study. Thus, we display in Figure~\ref{fig:butter30} only the higher latitudes above/below $+50\degr$/$-50\degr$. The gray regions indicate the polar magnetic-field reversal in the northern and southern hemisphere as described in \citet{Sun2015}. In the upper panel, we recognize structures at high latitudes above 70\degr. After the polar magnetic field reversal, no structures at latitudes above 75\degr\ are identified in the northern hemisphere. In the southern hemisphere, the polar magnetic field reversal occurred 16~months later, \textit{i.e.} in March 2014 \citep{Sun2015}. After the reversal, we detect only filaments that do not reach latitudes of $-75\degr$. These structures appear at around $-60\degr$.

The dash-to-the-Pole of polar crown filaments in the southern hemisphere started at approximately Carrington rotation 2115 and lasted until approximately Carrington rotation 2140 \citep{Xu2018}. The ChroTel observations start at the Carrington rotations 2122 and cover about $75\,\%$ of the dash-to-the-Pole. This allows us to estimate a rough migration rate for the PCFs for the southern hemisphere, which we compare with the migration rate given in \citet{Xu2018}. Cluster analyses was used to ensure that quiet-Sun filaments are not erroneously mixed with PCFs, which yields four clusters in the maximum phase (Figure~\ref{fig:butter30}). Clusters~I and~II (blue and green crosses) belong to the year 2012. They are clearly connected and represent the migration of the PCFs. Cluster~III (orange crosses) also belongs to the migration of the PCFs and is a continuation of the migration of the PCFs from Cluster~I and~II only separated by the gap of observations between 2012 and 2013. Cluster~IV (violet crosses) is excluded from the calculation of the migration rate, because it is clearly separated from cluster~III and represents ordinary quiet-Sun filaments. In order to verify the exclusion of Cluster~IV, we derived the distance of all data points to the line of best fit. All points from Cluster~IV have a significantly larger distance of more than $10\degr$ per rotation than the other detected PCFs of the other clusters. Ordinary quiet-Sun filaments are also excluded from the calculation of the migration rate in \citet{Xu2018} for Solar Cycles 21\,--\,23, as well. The derived migration rate from Clusters I\,--\,III is $0.79\degr\pm0.11\degr$ per rotation. An estimate of the migration rate for PCFs in the northern hemisphere is not possible based on ChroTel data due to the lack of observations in this phase.

\begin{figure} 
\centerline{
\includegraphics[width=1.0\textwidth]{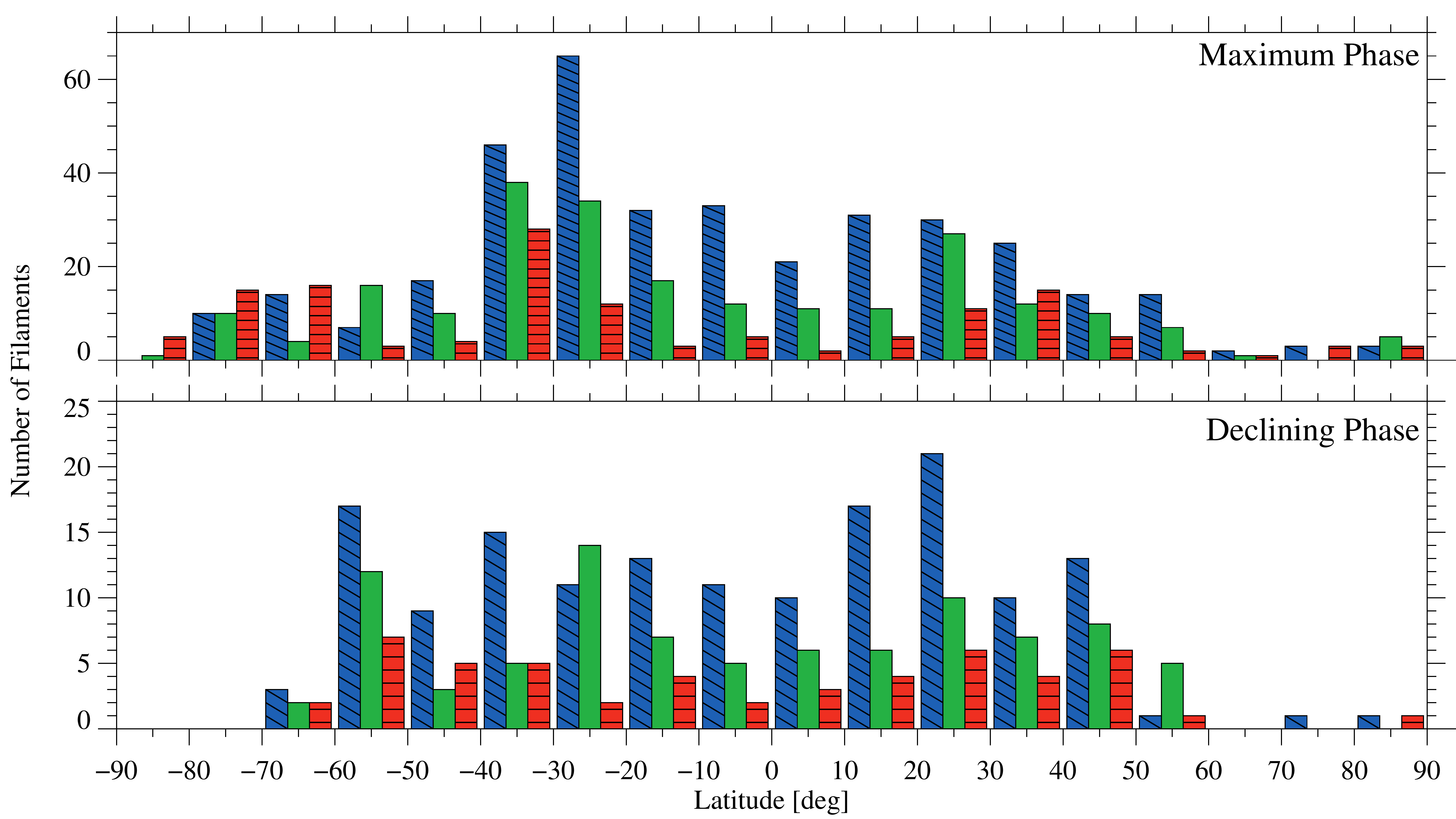}}
\caption{Distribution of filaments per area bin for the maximum (\textit{top}) and declining (\textit{bottom}) phase. The first area bin (\textit{filled blue with diagonal lines}) refers to filaments with an area smaller than 1000\,Mm$^2$, the second bin (\textit{filled green}) comprises areas between 1000\,Mm$^2$ and 2500\,Mm$^2$, and the third bin (\textit{filled red with horizontal lines}) contains areas larger than 2500\,Mm$^2$.}
\label{fig:harea}
\end{figure}

\subsection{Size Distribution of Filaments in Latitude}

Blob analysis facilitates extracting the area in pixels of detected filaments. The grid in the synoptic maps has an equidistant spacing between $\pm90\degr$ along the ordinate. Along the abscissa, each pixel corresponds to 0.1\degr. Thus, we calculated the area in megameters-squared and differentiated the filaments in three categories: i) $<$1000\,Mm$^2$, ii) 1000\,Mm$^2$\,--\,2500\,Mm$^2$, and iii) $>$2500\,Mm$^2$. In Figure~\ref{fig:harea}, we display a histogram of the three different size categories as a function of latitude. In the maximum phase (top panel in Figure~\ref{fig:harea}), the distribution across latitudes is imbalanced between the northern and southern hemisphere, \textit{i.e.} the northern hemisphere is dominated by the small and mid-sized filaments. The few detected filaments at high latitudes in the northern hemisphere belong to Category ii). In the southern hemisphere, many large filaments are present. In the declining phase (bottom panel Figure~\ref{fig:harea}), most of the detected filaments are smaller in both the northern and southern hemisphere, and only a few large filaments appear without any clear latitude dependence.

\subsection{Tilt Angle Distribution of Filaments}

\begin{figure} 
\centerline{
\includegraphics[width=1.0\textwidth]{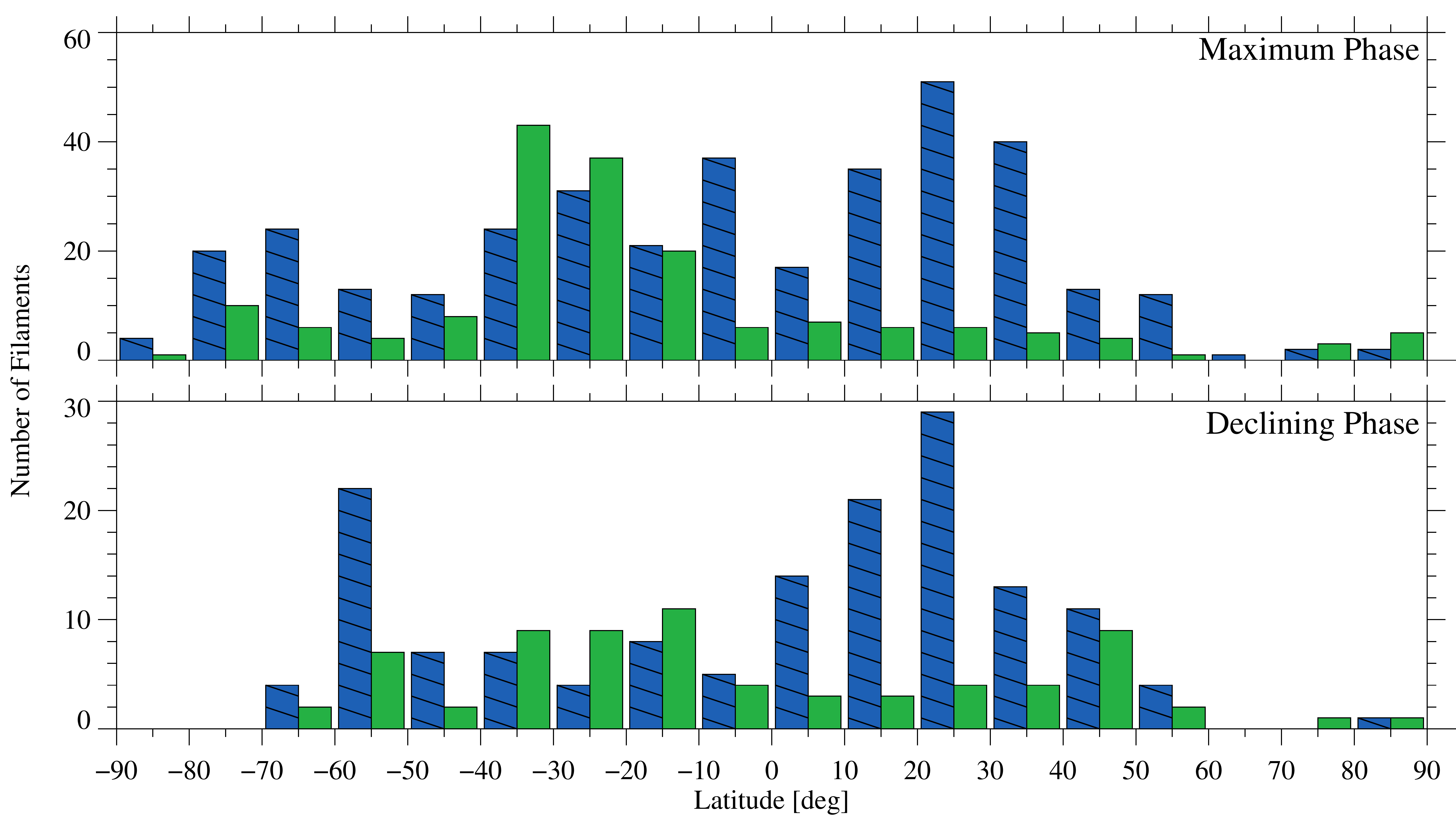}}
\caption{Distribution of filament tilt angles, which are either directed to the NW/SE (\textit{filled blue with diagonal lines}) or NE/SW (\textit{green}) direction for the maximum (\textit{top}) and declining (\textit{bottom}) phase.}
\label{fig:htheta}
\end{figure}

Fitting an ellipse to the detected filament yields the approximate orientation of the filaments relative to the Equator. We present in Figure~\ref{fig:htheta} the orientation of the detected filaments for different latitudes and for the maximum and declining phase. Essentially, we sorted the filaments according to their orientation in two categories: i) North-West (NW) or South-East (SE) orientation and ii) North-East (NE) or South-West (SW) orientation. According to Joy's Law the majority of active regions, and filaments as well, are orientated in the NW/SE direction in the northern hemisphere and in the NE/SW direction in the southern hemisphere \citep{Cameron2017}. In the maximum phase of Solar Cycle~24, the same relationship holds true for ChroTel data. Most of the detected filaments in the southern hemisphere have a NE/SW orientation, and in the northern hemisphere most of the objects have a NW/SE orientation. However, in the higher latitudes of the southern hemisphere, the NW/SE orientation is dominating. In the declining phase, the aforementioned relationship is again confirmed by the histogram for the northern hemisphere, whereas in the southern hemisphere, the relationship remains ambiguous. At mid-latitudes, both orientations turn up with the same frequency for bins between $-10\degr$ and $-40\degr$. However, the NW/SE orientation is more common in the bin $-60\degr$ to $-50\degr$.

\subsection{Distribution of H$\alpha$ and \mbox{He\,\textsc{i}} Line-Core Intensities}

\begin{figure} 
\centerline{
\includegraphics[width=1\textwidth]{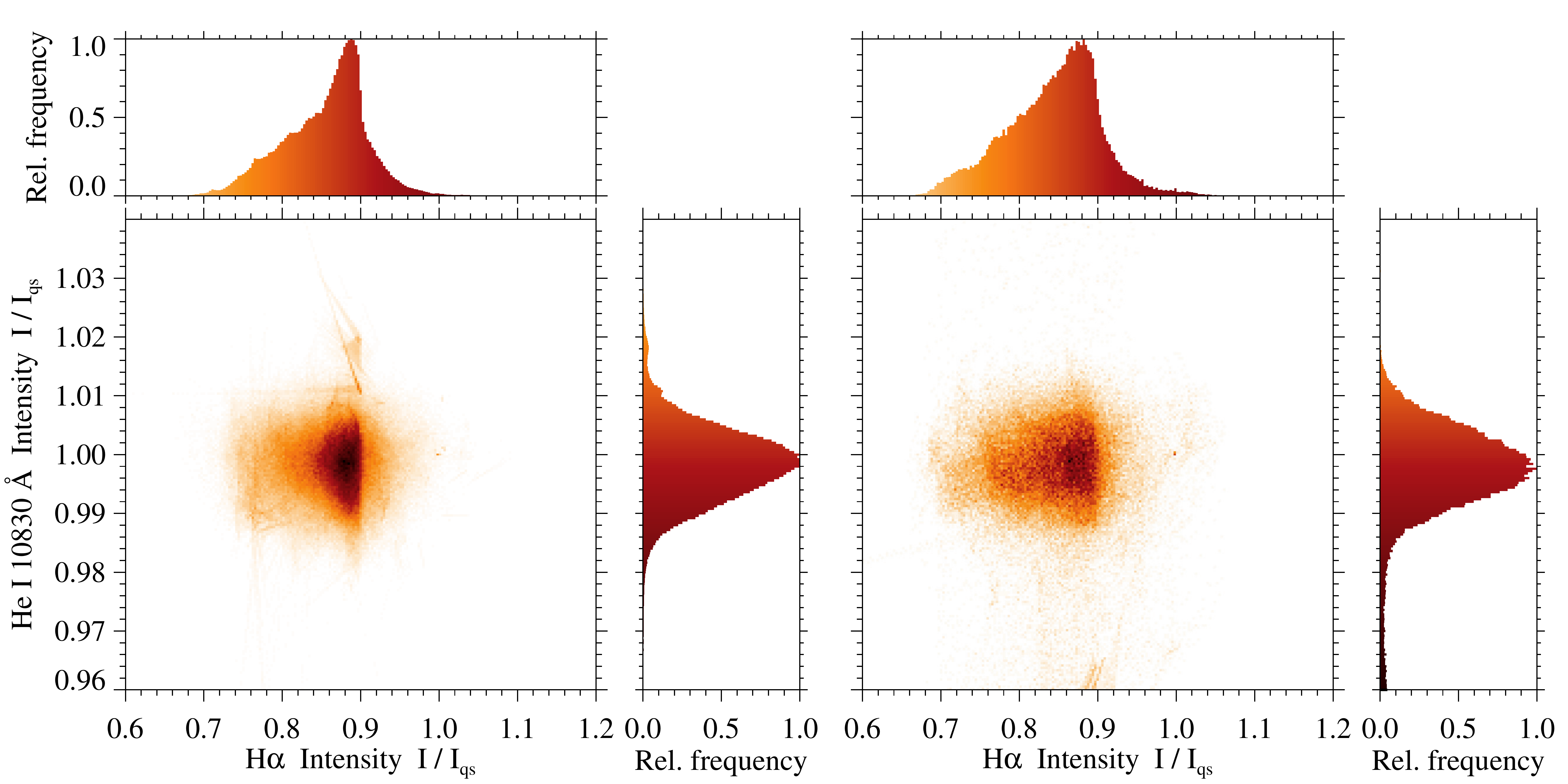}}
\caption{Scatter plot of H$\alpha$ and \mbox{He\,\textsc{i}} line-core intensities [$I/I_\mathrm{qs}$] derived from synoptic maps covering the maximum (\textit{left}) and declining (\textit{right}) phase for PCFs. In addition, we provide the respective histograms of the line-core intensities for each spectral line and phase of the cycle.}
\label{fig:scatter_part01}
\end{figure}

Figure~\ref{fig:scatter_part01} shows scatter plots of H$\alpha$ and \mbox{He\,\textsc{i}} line-core intensities for PCFs covering the maximum and declining phase, respectively. In addition, the individual histograms of the line-core intensities for H$\alpha$ and \mbox{He\,\textsc{i}} are provided. The images are normalized with respect to the mean quiet-Sun intensity [$I_\mathrm{qs}$], which was derived by creating a mask, which excluded dark and bright regions and computing the mean of the remaining values. The typical line-core intensity of the H$\alpha$ line is 16\,\% and for \mbox{He\,\textsc{i}} it is about 90\,\% of the continuum intensity in the quiet-Sun. Taking into account the FWHM of 0.5\,\AA\ of the H$\alpha$ Lyot filter, we would end up with a line-core intensity of 19\,\% for H$\alpha$ with respect to the continuum intensity. The values were obtained from a disk-integrated reference spectrum taken with the Kitt Peak Fourier Transform Spectral \citep[FTS:][]{Brault1985} atlas.

Comparing the histograms in the maximum phase and declining phase (Figure~\ref{fig:scatter_part01}) for PCFs, the H$\alpha$ histograms show that they cover a broader intensity range compared to the \mbox{He\,\textsc{i}} histograms but with a steeper negative slope toward higher intensities. The \mbox{He\,\textsc{i}} histograms show a Gaussian distribution and are more symmetric. Therefore, we fitted Gaussians to the \mbox{He\,\textsc{i}} histograms and derived a FWHM of 0.014 for both maximum and declining phase.

The two-dimensional histogram of the H$\alpha$ and \mbox{He\,\textsc{i}} line-core intensities of the maximum phase shows random structures at large \mbox{He\,\textsc{i}} intensities. This is likely caused by artifacts related to strong geometric distortions at very high latitudes close to the Poles. Another effect is introduced by the mask, which was derived from the H$\alpha$ images during the blob analysis. However, the filaments in \mbox{He\,\textsc{i}} are smaller than in H$\alpha$. Thus, structures of the surrounding quiet-Sun are included in the \mbox{He\,\textsc{i}} histogram. 

The histogram in H$\alpha$ for the declining phase differs from the one for the maximum phase. The slope toward the lower intensities is less steep and almost linear. This may be caused by the absence of erroneously detected structures close to the Pole. The slope toward brighter structures is instead much steeper, which can be also seen in the two-dimensional histogram. In total, there are fewer PCFs at higher latitudes in the declining phase. Only in the \mbox{He\,\textsc{i}} histogram, we recognize that the automatic detection found structures at lower \mbox{He\,\textsc{i}} intensities, which could be caused by geometric distortions close to the Pole.

%
\section{Discussion}\label{s:disc} 
%

We presented an overview of the ChroTel data set for the years 2012\,--\,2018 and created synoptic maps for each of the 90 Carrington rotation periods. With morphological image-processing tools, such as blob analysis, we automatically extracted dark objects, where we concentrated on elongated, large-scale objects, \textit{i.e.} various types of filaments. We carried out a statistical analysis and presented initial results based on the catalogue of detected filaments. 

In the first part of the analysis, we focused on the locations of the filaments as a function of time (Figure~\ref{fig:supersynopt}). The observations included the time of the polar magnetic field reversal for Solar Cycle~24 in both northern and southern hemispheres (Figure~\ref{fig:butter30}). We identified PCFs close to the Poles before the reversal, and we established their absence after the reversal. Up until the end of 2018, very few filaments were detected at high latitudes. For the southern hemisphere, we derived the migration rate of PCFs in Solar Cycle~24, \textit{i.e.} a rate of $0.79\degr\pm0.11\degr$ per rotation. The study of \citet{Xu2018} gives a rate of $0.63\degr\pm0.08\degr$ per rotation, which is smaller than our value. However, we have to take into account that our observations started after PCFs began migrating toward the Pole. Therefore, our estimate of the migration rate is based on incomplete data. Additional observations will increase the reliability of the migration rate. In addition, Figure~\ref{fig:butter30} illustrates the asymmetry of the northern and southern hemisphere in Solar Cycle~24, which was reported by \citet{Svalgaard2013} and \cite{Sun2015} for this cycle. Yet, an asymmetric behavior has been known since \citet{Spoerer1889} who studied the appearance of sunspots in both hemispheres. 

With upcoming observations in the next years, we can continue monitoring the propagation of high-latitude filaments toward the Pole, starting  after the solar minimum \citep{Hao2015, Xu2018}. This behavior of PCFs was expected and studied over several decades. Nonetheless, our work demonstrates that the ChroTel data set can be used for the analysis of the cyclic behavior of PCFs and their ``dash-to-the-Pole''. However, the ChroTel data set is relatively small compared to other full-disk H$\alpha$ surveys, \textit{i.e.} those of BBSO, KSO, or the Global Oscillation Network Group \citep[GONG: ][]{Harvey1996}, but it can complement the data sets of the other surveys. Similar to ChroTel, the full-disk telescopes at BBSO and KSO have a ten-centimeter aperture. The Lyot-type filter at BBSO has a bandpass of 0.25\,\AA\ and that of KSO a bandpass of 0.7\,\AA, whereas ChroTel's bandpass is 0.5\,\AA. All three surveys utilize a 2048$\times$2048-pixel detector size and contain images with a cadence as short as one minute. In addition, ChroTel provides \mbox{Ca\,\textsc{ii}\,K} $\lambda$3933\,\AA\ and \mbox{He\,\textsc{i}}~$\lambda$10830\,\AA\ filtergrams, whereby the \mbox{He\,\textsc{i}} spectroscopic data facilitate obtaining Dopplergrams. Full-disk data at KSO also include \mbox{Ca\,\textsc{ii}\,K} and continuum images. For filament research, which is the focus of our study, only  H$\alpha$ and \mbox{He\,\textsc{i}} filtergrams are relevant.  All instruments are ground-based facilities and are exposed to adverse weather conditions and periods of poor or mediocre seeing. Therefore, the H$\alpha$ data of all three surveys complement each other very well and partly bridge the day--night cycle for continuously monitoring solar activity.

Blob analysis provided the area for the detected filaments. We divided the data set chronologically according to the maximum (2012\,--\,2014, 528 observing days) and declining phase (2015\,--\,2018, 434 observing days) of Solar Cycle~24. In the maximum phase, the number of detected filaments is much larger than for the declining phase (Figure~\ref{fig:hnumber}), which is related to higher levels of solar activity but also related to the better coverage with ChroTel observations in this phase. For both phases, small filaments are more common at mid-latitudes below/above $+50\degr$/$-50\degr$ than filaments with an area greater than 1000\,Mm$^2$ (Figure~\ref{fig:harea}). In the maximum phase, we detected in general more filaments in the southern hemisphere between $-50\degr$ and $-30\degr$ as compared to the northern hemisphere or around the Equator. On the other hand, we find more large filaments ($>\!\!2500$\,Mm$^2$) at high latitudes as compared to small filaments. In addition, the number of filaments at high latitudes is smaller compared to mid-latitudes. 

\citet{Hao2015} compare the area of filaments for Solar Cycles~23 and~24 between 1996 and 2013 based on BBSO data. Both cycles show a symmetric increase of the number of filaments between $\pm40\degr$ and $\pm20\degr$ and a few filaments close to the Equator and the Poles. In our data set, we do not cover the activity maximum in the northern hemisphere, but only the maximum in the southern hemisphere, which explains the asymmetric histogram obtained from the ChroTel observations. In polar regions, \citet{Hao2015} did not find many PCFs with large areas. The automatized detection method applied to the ChroTel data found proportionally more filaments with larger areas in these regions. Nonetheless, large-scale filaments are not unusual close to the Poles and may be attributed to the later start of ChroTel observations during Solar Cycle~24 so that a number of larger filaments is better seen in our data set because of the smaller total number of detected filaments. 

The prevailing orientation of filaments in the northern (southern) hemisphere is north-west (north-east) \citep{Tlatov2016, Cameron2017}. This was confirmed with the ChroTel observations for both maximum and declining phase of Solar Cycle~24 at mid-latitudes between $\pm20\degr$ and $\pm50\degr$. In polar regions, the orientation seems to be reversed with a NE/SW orientation. The few detected filaments in the declining phase make it difficult to recognize a dominant orientation in the southern hemisphere, but in the northern hemisphere, a clear dominance of the NW/SE orientation is visible in the histogram. The studies of \citet{Hao2015} and \citet{Tlatov2016} confirm the dominant orientation of mid-latitude filaments taking into account more solar-cycle observations. \citet{Tlatov2016} observe a negative tilt angle for the polar regions, which is also visible in the ChroTel data of the maximum and declining phase. 

Finally, we inspected the line-core intensity distribution of filaments in H$\alpha$ and compared it to that of \mbox{He\,\textsc{i}} line-core intensities for the same regions. The distribution in H$\alpha$ is broader and resembles a bimodal distribution with a strong declining slope toward high intensities and a shallow slope toward low intensities, in both maximum and declining phase. The \mbox{He\,\textsc{i}} distribution is well-represented by a Gaussian distribution. We have to take into account that high-latitude filaments are located between $\pm50\degr$ and $\pm70\degr$. In addition, we recognize either lower or higher intensities, which do not belong to the overall intensity distribution in \mbox{He\,\textsc{i}}, which indicates that the filaments in \mbox{He\,\textsc{i}} have a smaller extent than in H$\alpha$. Thus, intensities of the surrounding quiet-Sun are erroneously included in the intensity distribution.

The study of \citet{Xu2018} focused on the location of filaments in the BBSO and KSO data for Solar Cycles 21\,--\,24. The ChroTel data set complements the data of BBSO and KSO. Information about tilt angles and size distribution is not yet available for BBSO and KSO data of Solar Cycle 24, which would be helpful for a more robust determination of PCF properties. Additional data of other H$\alpha$ surveys, \textit{e.g.} \textit{Kodaikanal Observatory} \citep{Chatterjee2017}, would allow us to cover more than a century of filament observations. An automatic data extraction approach with machine learning tools would facilitate an analysis of the lifetime and proper motions of the filaments. Space-based observations with, \textit{e.g.} the \textit{Solar Dynamics Observatory} \citep[SDO:][]{Pesnell2012}, would complement such a study with observations of the upper chromosphere, transition region, and corona.

%
\section{Conclusions}\label{s:conc} 
%

In this work, we demonstrated the value of the small-aperture synoptic solar telescopes such as the ten-cm aperture ChroTel. In 2012, the telescope started to take data on a regular basis, covering the maximum and declining phase of Solar Cycle~24. In total, it observed on almost 1000 days in seven years and recorded chromospheric images in three strong chromospheric absorption lines: H$\alpha$, \mbox{Ca\,\textsc{ii}\,K}, and near-infrared \mbox{He\,\textsc{i}}, whereby the \mbox{He\,\textsc{i}} line was scanned at seven different wavelength positions around the line core. We presented the image reduction pipeline, including limb-darkening correction, a non-uniform brightness correction with Zernike polynomials, and a geometric stretching to an equidistant graticule, which finally led to synoptic maps in all wavelengths. In this work, we concentrated on the automated extraction of filaments and their statistical analysis based on the H$\alpha$ data, \textit{i.e.} the number of filaments, their location, area, and tilt angle. Furthermore, we compared the line-core intensity distribution of filaments in H$\alpha$ with \mbox{He\,\textsc{i}} line-core intensities extracted from synoptic maps. Finally, we determined the appearance and disappearance of PCFs around the time of the polar magnetic field reversal, which enabled us to study the long-term evolution of PCFs in more detail. Other typical properties of filaments such as the tilt angle were derived from ChroTel observations. Hence, ChroTel can complement larger surveys, \textit{i.e.} those of BBSO, KSO, GONG, and many others \citep{Pevtsov2016}. Nonetheless, in contrast to the aforementioned observatories, ChroTel only started observations during the maximum of Solar Cycle~24 and has consequently a relatively smaller database. A significant advantage of ChroTel is that it observes the Sun regularly every three minutes in the three different wavelengths. Furthermore, high-temporal-resolution observations with a one-minute cadence are possible in H$\alpha$, which facilitates detailed studies of selected PCFs. 

The automated detection of filaments with morphological image processing tools, still contains false detections of small dark structures, which may not be related to filaments. We eliminated these structures in the present study with a rough threshold. Non-contiguous parts of filaments or small-scale active region filaments are also sorted out. In a forthcoming study, detection and recognition of filaments will be improved and automated with object detection methods based on neural networks. Thus, filaments can be analyzed in more detail revealing their physical properties, \textit{e.g.} class membership, length, and width. Furthermore, the rapid cadence of ChroTel observations enables us to study the temporal evolution of individual filaments and examining their horizontal proper motions. One future goal is to search for counter-streaming flows, which are ubiquitously present in all kinds of filaments. The high quality of the ChroTel data set facilitates the analysis of a large number of filaments in H$\alpha$ but also in \mbox{He\,\textsc{i}}. The latter provides line-of-sight velocities, which were not exploited so far. Structures such as plage regions appear bright in H$\alpha$ but dark in \mbox{He\,\textsc{i}}. The different morphologic and photometric properties of solar objects in the three wavelength bands of ChroTel are an excellent starting point for object tracking, identification, and classification using machine learning techniques and novel neural network routines.


\begin{acks}
The \textit{Chromospheric Telescope}
(ChroTel) is operated by the Leibniz Institute for Solar Physics (KIS) in Freiburg, Germany, at the Spanish Observatorio del Teide on Tenerife (Spain). The ChroTel filtergraph was developed by KIS in cooperation with the High Altitude Observatory (HAO) in Boulder, Colorado. This study was supported by grant DE~787/5-1 of the Deutsche Forschungsgemeinschaft (DFG) and by the European Commission's Horizon 2020 Program under grant agreements 824064 (ESCAPE -- European Science Cluster of Astronomy \& Particle Physics ESFRI Research Infrastructures) and 824135 (SOLARNET -- Integrating High Resolution Solar Physics). A. Diercke thanks Christoph Kuckein, Stefan Hofmeister, and Ioannis Kontogiannis for their helpful comments. The authors thank the reviewer for constructive criticism and helpful suggestions improving the manuscript.
\medskip

\noindent\textbf{Disclosure of Potential Conflicts of Interest}$\quad$ 
The authors declare that they have no conflicts of interest. 
\end{acks}

%
%


\end{article} 
\end{document}